\documentclass[%
reprint,
showpacs,preprintnumbers,
amsmath,amssymb,
aps,
floatfix,
prb,
]{revtex4-1}
\usepackage{amsmath}
\usepackage{graphicx}% Include figure files
\usepackage{dcolumn}% Align table columns on decimal point
\usepackage{csquotes}
\usepackage{bm}% bold math
\usepackage{float}
\usepackage{color}
\usepackage[colorlinks=true, allcolors=blue]{hyperref}
\usepackage{xpatch}
\usepackage[version=4]{mhchem}
\usepackage{pdfpages}

\xpretocmd{\eqref}{Eq.~}{}{}
\graphicspath{ {./figures/} }

\makeatletter
\AtBeginDocument{\let\LS@rot\@undefined}
\makeatother

\begin{document}

%\preprint{APS/123-QED}
\title{Kinetics of lithium electrodeposition and stripping}

%%%%%%%%%%%%%%%%%%%%%%%%%AUTHORS%%%%%%%%%%%%%%%%%%%%%%%%%%%%%%%%%%%
\author{Shashank Sripad}
\thanks{These two authors contributed equally}
\affiliation{%
Department of Mechanical Engineering, Carnegie Mellon University, Pittsburgh, Pennsylvania 15213
}

\author{Daniel Korff}
\thanks{These two authors contributed equally}
\affiliation{Department of Mechanical Engineering, Colorado School of Mines, Golden, Colorado, 80401}

\author{Steven C. DeCaluwe}
\affiliation{Department of Mechanical Engineering, Colorado School of Mines, Golden, Colorado, 80401}

\author{Venkatasubramanian Viswanathan}%
\email{venkvis@cmu.edu}
\affiliation{%
Department of Mechanical Engineering, Carnegie Mellon University, Pittsburgh, Pennsylvania 15213
}

\date{\today}% It is always \today, today,
             %  but any date may be explicitly specified

\begin{abstract}
Electrodeposition and stripping are fundamental electrochemical processes for metals and have gained importance in rechargeable Li-ion batteries due to lithium metal electrodes.  The electrode kinetics associated with lithium metal electrodeposition and stripping is crucial in determining performance at fast discharge and charge which is important for electric vertical take-off and landing (eVTOL) aircraft and electric vehicles (EV).  In this work, we show the use of Marcus-Hush-Chidsey (MHC) kinetics to accurately predict the Tafel curve data from the work of Boyle et al. [ACS Energy Lett. 2020, 5, 3, 701] We discuss the differences in predictions of reorganization energies from the Marcus-Hush and the MHC models for lithium metal electrodes in four solvents. The MHC kinetic model is implemented and open-sourced within \textit{Cantera}. Using the reaction kinetic model in a pseudo-2D battery model with a lithium anode paired with a \ce{LiFePO4} cathode, we show the importance of accounting for the MHC kinetics and compare it to the use of Butler-Volmer (BV) and Marcus kinetic models.  We find that significant deviation in the overpotentials associated with reaction kinetics for the two different rate laws for conditions of fast discharge and charge relevant for eVTOL and EV respectively.
\end{abstract}

\maketitle
%\tableofcontents

\section{Introduction}
There is a need for improved energy density and specific energy for electrifying transportation and aviation.\cite{sripad2017performance,bills2020performance}  Lithium metal electrodes are a promising avenue to improve the energy density to meet these requirements.\cite{Albertus2017} Reactions and kinetics at electrode-electrolyte interfaces limit performance of lithium metal electrodes. For a lithium metal electrode, during charge, lithium ions from the electrolyte (solid or liquid) deposit at the metal electrode, popularly known as electrodeposition, while during discharge, lithium metal is oxidized into lithium ions, through a process known as stripping.\cite{Krauskopf2020}

Fast charging is an important requirement for electric vehicles,\cite{Albertus2017} while fast discharge is an important requirement for the take-off and landing segment of electric vertical take-off and landing aircraft.\cite{fredericks2018performance}  The morphological instabilities related to both fast discharge and charge have been well-documented leading to pitting and dendrite formation,\cite{Wood2016,Kasemchainan2019} thereby leading to poor cycle life.

Despite the enormous attention paid to understanding morphology, the kinetic rate behavior at the lithium metal electrode-electrolyte interface has received much less attention.  Typical lithium metal models usually invoke Butler-Volmer kinetics,\cite{doyle1993modeling} despite wide-recognition from the electrodeposition community that kinetic laws beyond Butler-Volmer are needed to describe fast discharge and charge conditions.  Recently, Boyle \textit{et al}.,\cite{boyle2020transient} measured lithium electrodeposition and stripping kinetics using transient voltammetry.  Based on the measured data, they observed significant deviations from Butler-Volmer kinetics and argued for the use of Marcus-Hush kinetics implemented as a low overpotential approximation of the Marcus-Hush-Chidsey formalism.  However, the low overpotential approximation breaks down for fast charge and discharge regimes.

In this work, building on the work of Boyle \textit{et al},\cite{boyle2020transient} we show that the Marcus-Hush-Chidsey (MHC) model implemented as a uniformly valid closed form approximation developed by Zeng and coworkers\cite{zeng2014simple} agrees with the lithium electrodeposition and stripping data.\cite{boyle2020transient}
We incorporate the MHC kinetic model it into a pseudo-2D model and demonstrate for constant current fast charge and discharge applications up to current rates of 12C, and fast discharge eVTOL missions with discharge rates of up to 10C. We find that the MHC kinetic law is important to track performance under these high rate operating conditions.  In order to enable wide-spread use, we open-source the kinetic model for lithium metal into Cantera\cite{cantera}.  We believe that future studies with lithium metal electrodes should utilize this formalism.

\section{Methods}
\subsection{Electron transfer kinetics}\label{methodskinetics}
There are several approaches to modeling electron transfer kinetics,\cite{bard2001fundamentals} beginning in the 1930s with Butler-Volmer model,\cite{butler1932mechanism,erdey1930theory} microscopic electron transfer kinetics proposed by Marcus\cite{marcus1956theory} in the 1950s, kinetics of electrode processes examined by Hush,\cite{hush1958adiabatic,hush1999electron} and more recently, heterogeneous electron transfer explained using the Marcus-Hush-Chidsey\cite{chidsey1991free} kinetic model. Each of these kinetic models are presented with required modifications for their suitable interpretation in the context of electrochemical performance models of Li-ion batteries. Each kinetic model is presented for the following reaction, 
\begin{equation}
 \mathrm{Li^+ + e^- \rightleftharpoons Li}   \label{e1}
\end{equation}
The charge transfer kinetics at the interface of an electrode in electrochemical energy devices is often modeled using Butler-Volmer kinetics.\cite{bard2001fundamentals,newman2012electrochemical} Over the last several decades, the Butler-Volmer (BV) formalism has been used to model the kinetics at lithium metal anodes of Li-ion batteries.\cite{newman2012electrochemical,doyle1993modeling,Ramadesigan_2012,Monroe_2005,Srinivasan2020,hao2018mesoscale} BV formalism can be written as,
\begin{align}
    \mathrm{k_{red}^{BV}} &= \mathrm{k_0^{BV}.exp\bigg(\frac{-\alpha F}{RT}\eta\bigg)}\label{bv1}\\
    \mathrm{k_{ox}^{BV}} &= \mathrm{k_0^{BV}.exp\bigg(\frac{(1-\alpha) F}{RT}\eta\bigg)}\label{bv2}
\end{align}
where $\mathrm{k_{red}}$ and $\mathrm{k_{ox}}$ are the reaction rates for reduction and oxidation respectively, $\mathrm{k_0^{BV}}$ is the rate constant, $\mathrm{\eta}$ is the applied overpotential, and $\mathrm{\alpha}$ is transfer coefficient which represents the position of the transition state of the reaction.\cite{laborda2013asymmetric} The transfer coefficient, $\mathrm{\alpha}$, is a constant value which is generally 0.5 for single electron processes like the one shown in \eqref{e1}. The BV model posits a linear relationship between $\mathrm{ln(k_{red}^{BV}/k_0^{BV})}$ and the overpotential $\mathrm{\eta}$ with a slope of $\mathrm{-\alpha F/RT}$ which is independent of the potential. However, experiments by Sav{\'e}ant and Tessier\cite{saveant1975convolution} and many other studies show considerable deviations from linearity.\cite{bard2001fundamentals}  The constant transfer coefficient independent of potential is hence considered only a qualitative metric from BV theory and not universally applicable.\cite{bard2001fundamentals,laborda2013asymmetric} 

Around 65 years ago, Marcus developed a kinetic model for microscopic outer-sphere homogeneous electron transfer.\cite{marcus1956theory} The theory has been used to model kinetics in several chemical and biological systems.\cite{marcus1985electron} The rate expressions for the Marcus model can be written as,
\begin{equation}
    \mathrm{k_{red/ox}^M} = \mathrm{k_0^M.exp\bigg(-\frac{(\Delta G\pm\lambda)^2}{4\lambda.RT}\bigg)}\label{marcuseq}
\end{equation}
where $\mathrm{\Delta G}$ is the free energy change on reduction and $\mathrm{\lambda}$ is the reorganization energy, that required to reorganize the nuclear configuration of the reactants and solvent to the product state.\cite{bard2001fundamentals} One of the important predictions from Marcus theory is dependence of the transfer coefficient on the overpotential. The relevant derivations can be found elsewhere.\cite{marcus1956theory,marcus1965theory,bard2001fundamentals,saveant1975convolution} The potential dependence of $
\mathrm{\alpha}$ is expressed as,
\begin{equation}
    \mathrm{\alpha=\frac{1}{2}+\frac{F\eta}{4\lambda}} \label{alpha}
\end{equation}
which can be accommodated into \eqref{bv1} and \eqref{bv2}, however, classical BV theory treats $\mathrm{\alpha}$ as a constant.\cite{bard2001fundamentals} The use of a potential dependent $\mathrm{\alpha}$ was shown by Sav{\'e}ant and Tessier\cite{saveant1975convolution} and has also been used to explain recent data on lithium electrodeposition and stripping by Boyle \textit{et al}.\cite{boyle2020transient}

The Marcus kinetic model is developed for homogeneous electron transfer reactions where the two species involved in the reaction are dissolved in a solution. However, for applications like metal electrodes, the electrode charge transfer reaction is heterogeneous,\cite{henstridge2012marcus,laborda2013asymmetric} i.e. between a metal electrode and a species dissolved in a solution. For heterogeneous electron transfer with metal electrodes, the kinetics of the reaction will depend on electrons from different energy levels in the conduction band of the metal electrode.\cite{marcus1965theory} The dependence of reaction rates on the ``electronic energy levels in metal" was noted by Marcus in 1965,\cite{marcus1965theory} and the incorporation of Fermi-Dirac statistics and its validation was shown by Chidsey in 1991,\cite{chidsey1991free} with a gold electrode and an electroactive ferrocene group.\cite{henstridge2012marcus,laborda2013asymmetric,zeng2014simple} The rate equation used by Chidsey is generally referred to as the Marcus-Hush-Chidsey model,
\begin{equation}
    \mathrm{k_{red/ox}^{MHC}} = \mathrm{A.\int^\infty_{-\infty}exp\bigg(-\frac{(x-\lambda\mp F\eta)^2}{4\lambda.RT}\bigg).\frac{dx}{1+exp(x/RT)}} \label{mhceq}
\end{equation}
where $\mathrm{A}$ is a pre-exponential factor which accounts for the density of metallic states and the electronic coupling strength.\cite{zeng2014simple,laborda2013asymmetric} Compton and coworkers explain that the assumption of $\mathrm{A}$ being independent of potential has generally proven to be reasonable with experiments.\cite{nissim2012electrode,laborda2013asymmetric,chidsey1991free} Further, for the case of lithium, the density of states is relatively flat based on density functional theory calculations and the validity of this assumption is tackled in an accompanying paper for this Special Issue.\cite{kurchin2020} The activation energy in \eqref{marcuseq} is recast using the energy of the electron described as $\mathrm{x}$, defined relative to the Fermi level.\cite{henstridge2012marcus,zeng2014simple,laborda2013asymmetric} 

The net current density from the reaction, $\mathrm{j}$, can be written as, $\mathrm{j=j_0.(k_{red}-k_{ox})}$, where $\mathrm{j_0}$ is the exchange current density. The net current for the BV model can be obtained using \eqref{bv1} and \eqref{bv2},
\begin{equation}
    \mathrm{j^{BV}=j_0^{BV}.(k_{red}^{BV}-k_{ox}^{BV})}\label{bvcurrent}
\end{equation}

For applying Marcus theory, as noted previously, Boyle et al.,\cite{boyle2020transient} employed a potential dependent $\mathrm{\alpha}$\cite{marcus1956theory,marcus1965theory,bard2001fundamentals,saveant1975convolution} as shown in \eqref{alpha} within a BV current expression as shown in \eqref{bvcurrent}. This model is referred to as the Marcus-Hush (M-H) kinetic model. The net current of the M-H model $\mathrm{j^{M-H}}$ would have its exchange current density $\mathrm{j^{M-H}_0}$ and follow the functional form shown in \eqref{bvcurrent} along with $\mathrm{\alpha}$ described by \eqref{alpha}.

A closed form expression for the net current using the MHC model is required to avoid the evaluation of the MHC integral in \eqref{mhceq} which is cumbersome to implement in a battery model. Bazant and coworkers have developed a uniformly valid closed form approximation of the integral,\cite{zeng2014simple} where they obtain a net current density expression using dimensionless reorganization energy, $\mathrm{\lambda^\ast=\lambda/RT}$, and dimensionless overpotential $\mathrm{\eta^*=F.\eta/RT}$,
\begin{equation}
    \mathrm{j^{MHC} \approx S. \sqrt{\pi\lambda^*}.tanh\bigg(\frac{\eta^*}{2}\bigg).erfc\bigg(\frac{\lambda^*-\sqrt{1+\sqrt{\lambda^*}+\eta^{*2}}}{2\sqrt{\lambda^*}}\bigg)} \label{mhcapprox}
\end{equation}
where $\mathrm{S}$ is used as a scaling factor for calculating the current density and $\mathrm{erfc}$ is the complementary error function. The closed form approximation provides the exchange current density to be,\cite{zeng2014simple}
\begin{equation}
    \mathrm{j_0^{MHC} \approx S. \frac{\sqrt{\pi\lambda^*}}{2}.erfc\bigg(\frac{\lambda^*-\sqrt{1+\sqrt{\lambda^*}}}{2\sqrt{\lambda^*}}\bigg)} \label{jomhcapprox}
\end{equation}
which is a function of the reorganization energy alone. This approximation has been shown to be consistent with the numerical integration of \eqref{mhceq}, with the accuracy increasing with the overpotential.\cite{zeng2014simple} Given that the goal of this work is to examine the implications of predictions from different kinetic models for high power applications of lithium batteries, we find that \eqref{mhcapprox} provides the required flexibility and simplicity of the BV model, while preserving the high accuracy of the MHC kinetic model. 

We use transient voltammetry data from Boyle \textit{et al}.,\cite{boyle2020transient} collected using ultramicroelectrodes for $\mathrm{LiPF_6}$ in four solvents, (i) propylene carbonate (PC), (ii) diethyl carbonate (DEC), (iii) 1:1 by volume ethylene carbonate: diethyl carbonate (EC:DEC), and (iv) EC:DEC with 10\% fluoroethylene carbonate (EC:DEC w. 10\% FEC) to fit $\mathrm{\lambda}$'s which we will refer to as $\mathrm{\lambda^{M-H}}$ for the Marcus-Hush model and $\mathrm{\lambda^{MHC}}$ for the MHC model, respectively. The M-H model is implemented using \eqref{alpha} and the MHC model is implemented using \eqref{mhcapprox}. We also fit values for the exchange current densities $\mathrm{j^{M}_0}$ and $\mathrm{j^{MHC}_0}$ for the two kinetic models. The curve fitting was performed using the non-linear least squares method on MATLAB r2019b.

\subsection{Pseudo-2D fast-charge model}\label{p2dmodel}
While the fundamental charge-transfer kinetic models described above are of no doubt great theoretical interest, do they result in a meaningful performance difference at technologically relevant scales? As shown in the aforementioned works,\cite{zeng2014simple,laborda2013asymmetric} the models diverge only when the overpotential $\eta$ is significant, relative to the solvent reorganization energy $\lambda$. With the push toward extreme fast charge (XFC) rates to facilitate EV consumer adoption and fast discharge for eVTOLs, the probability of realizing suitably large overpotentials increases, and merits investigation.

To explore the impact of detailed charge transfer kinetics on battery performance, we employ here a physically-based pseudo-2D (P2D) model based on that developed by Newman \textit{et al.}\cite{doyle1993modeling}.  The model simulates the performance of a porous LiFePO$_4$ cathode, using porous electrode theory, paired with a dense lithium metal anode. We implement the three charge-transfer kinetic models as part of the open-source chemical kinetics software package Cantera\cite{cantera}, which is used to manage the thermo-kinetic calculations in a generalized manner.  

Despite the lower gravimetric capacity and electrical conductivity of LiFePO$_4$, it has several benefits, such as lower material cost (6.3 Wh US\$$^{-1}$) than other standard cathode materials \cite{howard2007LiPhosphates} while having a high theoretical capacity among lithium metal phosphates, which tend to be more stable and have a longer cycling and calendar life \cite{howard2007LiPhosphates}. Additionally, advances in LFP cathodes have improved the specific capacity and electrical conductivity of the material \cite{wang2005FeDoping, meng2019LFPdoping}, which, combined with favorable charge transfer kinetics at the electrode/electrolyte interface\cite{jow2012LFPkinetics} and high cyclability \cite{meng2019LFPdoping} make LFP a competitive cathode material for fast charge and high power applications \cite{zhou2011graphene, li2019uhcLFP}.

\subsubsection{State Variables} The solution variables tracked throughout the charge/discharge of the cell are:\\
in the LiFePO$_4$ phase:

\begin{itemize}
    \item $\mathrm{ X_{\text{Li,\,ca}} }$ the intercalation fraction in the cathode (-)
    \item $\mathrm{\Phi_\text{ca} }$ the electric potential of the cathode (V)
\end{itemize}

in the electrolyte phase:

\begin{itemize}
    \item $\mathrm{ X_{k,\text{elyte}} }$ the mole fraction of species \textit{k} in the electrolyte ($\mathrm{\text{kmol}_k ~ \text{kmol}^{-1}}$)
    \item $\mathrm{ \Phi_\text{elyte} }$ the electric potential of the electrolyte (V)
\end{itemize}

in the anode SEI:

\begin{itemize}
    \item $\mathrm{\Phi_{SEI/anode}}$ the electric potential in the SEI at the anode interface (V).
\end{itemize}

and in the lithium:

\begin{itemize}
    \item $\mathrm{ \Phi_\text{an} }$ the electric potential of the lithium (V)
\end{itemize}

\subsubsection{Model equations}
The governing equations are derived from physically based conservation equations of mass, species, and electrical charge. The equations, described below, represent a set of differential-algebraic equations, which are integrated temporally to simulate galvanostatic charge-discharge with a fixed external current $\mathrm{j_\text{ext}}$. While the P2D model equations are common and well-represented throughout the literature\cite{doyle1993modeling, colclasure2010, chiew2019, Murbach_2017}, we present here a brief overview of the governing equations.\\

{\bf Cathode Composition}. In the cathode, assuming spherical particles, diffusion only in the radial direction, and a constant diffusion coefficient $\mathrm{D_\text{Li,\,ca}}$ the governing equation for the intercalated lithium mole fraction is derived from conservation of mass and elements:

\begin{equation}
    \mathrm{\frac{\partial X_\text{Li,\,ca}}{\partial t} = \nabla \cdot (D_\text{Li,\,ca} \nabla X_\text{\rm Li,\,ca} + \frac{\dot{s}_\text{Li}A_\text{int}}{C_\text{ca}})}
\end{equation}\\

\noindent where $\mathrm{C_\text{ca}}$ is the total molar concentration of the cathode phase (mol m$^{-3}$), and the term $\mathrm{\dot{s}_\text{Li}}$ is a molar source term due to the charge transfer reaction at the LFP particle surfaces (mol m$^{-2}$ s$^{-1}$). $\mathrm{A_\text{int}}$ is the specific surface area of the electrode/electrolyte interface ($\mathrm{\text{m}_{\text{int}}^2~\text{m}^{-3}}$).  At the particle surface, it is evaluated via one of the three charge transfer kinetic models described above, with the relationship between molar production rate and the Faradaic current given by Faraday's law:

\begin{equation}
    \mathrm{\dot{s}_{k,\,elyte} = \nu_{k,\,Li}\frac{j_{Far}}{zF}},
\end{equation}
where $\mathrm{\nu_{k,\,Li}}$ is the stoichiometric coefficient for intercalated lithium in the cathode phase (+1, here) and z represents the moles of charge transferred per mole of reaction ($\mathrm{z=1}$ in this study). A similar relationship is used for the electrolyte species production rates, below. Lastly, $\dot{s}_\text{Li}A_\text{int}$  is equal to zero, internal to the particle.\\

{\bf Electrolyte Composition}. Conservation of mass and elements are also applied to derive the governing equation for the electrolyte species mole fractions:

\begin{equation}
    \mathrm{\frac{\partial X_{k,\text{elyte}}}{\partial t} = \frac{1}{C_\text{elyte} \varepsilon_\text{elyte}}\left(-\nabla N_{k,\,elyte} + \dot{s}_{k,\,{\rm elyte}} A_\text{int}\right)},
\end{equation}
where $\mathrm{C_\text{elyte}}$ is the molar concentration of the electrolyte ($\mathrm{\text{mol}~\text{m}^{-3}}$, assumed constant for the incompressible electrolyte), $\mathrm{\varepsilon_{\rm elyte}}$ is the electrolyte volume fraction, ${N_{k,\,\text{elyte}}}$ is the molar flux ($\mathrm{\text{mol}_k~\text{m}^{-2}~s^{-1}}$), calculated according to concentrated solution theory\cite{doyle1993modeling, Neidhardt_2012,NYMAN20086356}, and $\mathrm{\dot{s}_{k,\,\text{elyte}}}$ is the net production rate of electrolyte species k ($\mathrm{\text{mol}_k ~ \text{m}^{-2}_\text{int} ~ \text{s}^{-1}}$) due to heterogeneous reactions at the electrode/electrolyte interface.  Within the electrolyte separator, the heterogeneous term $\mathrm{\left(\dot{s}_k A_\text{int}\right)}$ is equal to zero.\\

{\bf Electric Potentials} Finally, governing equations for the electrode and electrolyte electric potentials are derived by application of charge conservation while concurrently assuming charge neutrality in bulk phase interiors. This yields an equation for the double-layer potential at all electrode-electrolyte interfaces $\mathrm{\Delta \Phi_\text{dl} = \Phi_\text{elyte} - \Phi_\text{ed}}$ (V, where `ed' $=$ `ca' (cathode) or `an' (anode)), and an algebraic constraint on the sum of all currents into the volume:
\begin{equation}
    \mathrm{\frac{\partial \big( \Phi_\text{elyte} - \Phi_\text{ed} \big)}{\partial t} = \frac{j_\text{dl}}{C_\text{dl,ed} A_\text{surf}}}
\end{equation}
and
\begin{equation}
    \mathrm{0 = \nabla j_\text{io} + \nabla j_\text{el}}.
\end{equation}
In these equations, $\mathrm{j_\text{dl}}$ is the double-layer current per unit volume at the electrode/electrolyte interface ($\mathrm{\text{A}~\text{m}^{-3}}$), $\mathrm{C_{\text{dl},\text{ed}}}$ is the double-layer capacitance, and $\mathrm{j_\text{io}}$ and $\mathrm{j_\text{el}}$ are the ionic and electronic current densities, respectively ($\mathrm{\text{A}~\text{m}^{-2}}$, both defined as positive for positive charge moving in the positive $y$ direction).  The double layer current is calculated by enforcing charge neutrality within the bulk of either the electrode or electrolyte phase (without loss of generality, we choose the electrode, here):
\begin{equation}
    \mathrm{0 = \nabla j_\text{el} + j_\text{Far} + j_\text{dl}},
\end{equation}
where $\mathrm{j_\text{Far}}$ is the Faradaic current at the electrode/electrolyte interface due to charge transfer reactions, per unit volume ($\mathrm{\text{A}~\text{m}^{-3}}$).  The interfacial currents $\mathrm{j_\text{Far}}$ and $\mathrm{j_\text{dl}}$ are both defined such that positive $\mathrm{j}$ represents positive charge entering the electrode and leaving the electrolyte.  Therefore, positive $\mathrm{j_\text{dl}}$ represents an increase in the positive charge on the electrolyte side of the charged double layer.

The handling of the double layer here represents one potential point of difference, relative to other P2D models. Some models\cite{colclasure2010, chiew2019} assume that the double layer responds infinitely quickly, such that $\mathrm{j_\text{dl}}$ is zero at all times.  This converts the problem to an additional algebraic constraint, such that the $\mathrm{\Phi_\text{elyte}}$ must always be set to that value which results in $\mathrm{j_\text{Far} = - \nabla j_\text{el}}$. Whereas the approach in this study results in a differential equation, and captures the dynamic response of the double layer during battery charge and discharge, which can have non-negligible impacts, particularly for XFC applications.

Lastly, although Boyle, et al. demonstrate that SEI dynamics do not significantly impact the electrode kinetics in their transient voltammetry experiments, ~\cite{boyle2020transient} SEI will certainly form and influence performance during charge-discharge. We estimate here the impact of an SEI with constant resistance (0.032 $\Omega-\text{m}^2$ is used, here~\cite{colclasure2011sei}).  This imposes an additional geometric constraint, in that the current across the SEI (modeled here as Ohmic in nature) must equal the Faradaic current at the SEI/anode interface.  In other words, the model determines the SEI electric potential at the anode interface ($\Phi_{\rm SEI/anode}$) such that $\mathrm{j_{Far} = j_{SEI}}$).  The charge transfer kinetics use the potential difference $\Phi_{\rm SEI/anode} - \Phi_{\rm anode}$ to calculate the overpotential $\eta = \Delta \Phi - \Delta \Phi^{\rm eq}$.\\

{\bf Model and Simulation Parameters} Model parameters were taken from previous experimental and modeling studies for Li--LiFePO$_4$/C.  The P2D simulations are presented here to demonstrate the impact of charge transfer kinetics within a relevant application, not as a fully validated, predictive modeling study. As such, model parameters were adapted to qualitatively match rate capability data from recent literature,~\cite{Tang2019go_lfp} but have not been fit to the data.  This is left as an extension for a future study.

Simulations were run for two conditions: (i) galvanostatic charge-discharge at C-rates ranging from 1C to 15C, and (ii) simulated eVTOL flight, with a 1-minute take-off and landing segments at 10C, and a 5-minute cruise at 5C.  The source code for the model, including instructions for installation and running, as well as all parameter values, is made available via GitHub\cite{p2dCode}.  Although we plan to incorporate the charge transfer kinetics into an official future release of Cantera, for the time being the required functionality can be obtained by using a forked version of the chemical thermo-kinetics software, published as a GitHub repo\cite{canteraFork}.

\section{Results}
The transient voltammetry data for lithium electrodeposition and stripping from Boyle \textit{et al}.,\cite{boyle2020transient} is evaluated for the M-H model and the MHC model described in Section \ref{methodskinetics} to find the exchange current densities, $\mathrm{j_0}$'s, and the reorganization energies, $\mathrm{\lambda}$'s. We compare the Marcus-Hush (M-H) kinetic model proposed by Boyle \textit{et al}., and our implementation of the MHC model along with the BV model. Each of these are evaluated in the context of using these kinetic models within lithium battery models to supplant the existing BV models. We find that the MHC model is more appropriate to use within electrochemical battery models. We compare the predictions from the P2D model in fast charge and discharge conditions for the BV, M-H and MHC kinetic models.

\subsection{Electron transfer kinetics}
The best fit exchange current densities for the M-H model ($\mathrm{j_0^{M-H}}$), the MHC model ($\mathrm{j_0^{MHC}}$), along with the reported linear fit values ($\mathrm{j_0^{\textit{L.fit}}}$) from Boyle \textit{et al}.,\cite{boyle2020transient} are shown in Table \ref{table:jofit}. The values of exchange current density fit for the M-H and MHC model in this work agree closely with each other. However, the $\mathrm{j_0^{\textit{L.fit}}}$ values are seen to be marginally higher than $\mathrm{j_0^{M-H}}$ and $\mathrm{j_0^{MHC}}$. The trends in exchange current density hold in each of the estimates, where the exchange current density of PC$<$DEC$<$ EC:DEC$<$EC:DEC w. 10\% FEC. As noted by Boyle \textit{et al}., the current density predicted the transient voltammetry data is not affected by the solid electrolyte interphase (SEI),\cite{boyle2020transient} and the reaction is in kinetic control.

\begin{table}[ht]
\centering
\vspace{10pt}
 \begin{tabular}{|p{3.3cm}|p{1.5cm}|p{1.5cm}|p{1.5cm}|} 
 \hline
 & & & \\[-0.5em]
 Solvent & $\mathrm{j_o^{M-H}(\frac{mA}{cm^2})}$ & $\mathrm{j_o^{MHC}(\frac{mA}{cm^2})}$ & $\mathrm{j_o^{\textit{L.fit}}(\frac{mA}{cm^2})}$ \\[0.3em]
  & & & Ref. [\citenum{boyle2020transient} ] \\
 \hline\hline
 PC& 1.9 & 1.9& 2.6\\
 \hline
 DEC& 2.2& 2.2& 3.7\\
 \hline
 EC:DEC& 8.8& 8.6& 10.4\\
 \hline
 EC:DEC w. 10\% FEC&14.5 & 13.8& 16.0\\
 \hline
\end{tabular}
\caption{The exchange current density values as obtained from the M-H model, $\mathrm{j_o^{M-H}}$, the MHC model, $\mathrm{j_o^{MHC}}$, and values reported by Boyle \textit{et al}., using a linear fit.}\label{table:jofit}
\end{table}

The best fit values for the reorganization energy for the M-H model ($\mathrm{\lambda^{M-H}}$), the MHC model ($\mathrm{\lambda^{MHC}}$) along with the reported M-H fit values from Boyle \textit{et al}.,\cite{boyle2020transient} are shown in Table \ref{table:lamfit}. We observe that the $\mathrm{\lambda^{M-H}}$ values of our study and $\mathrm{\lambda^{M-H\diamond}}$ from Boyle \textit{et al}., are similar, however, $\mathrm{\lambda^{MHC}}$ values are seen to be generally lower than $\mathrm{\lambda^{M-H}}$ values for each of the solvents. The relative values and the trend for the reorganization energy between the four solvents holds across M-H and MHC models. The calculation of $\mathrm{\lambda^{M-H}}$ and $\mathrm{\lambda^{M-H\diamond}}$ neglects the distribution of electrons, while $\mathrm{\lambda^{MHC}}$ calculated using \eqref{mhcapprox} implicitly accounts for it. While the exchange current density values of our M-H and MHC models agree closely, the reorganization energy values are consistently lower for MHC by about 0.12$\mathrm{\pm}$0.1 eV. 

\begin{table}[ht]
\centering
\vspace{10pt}
 \begin{tabular}{|p{3.3cm}|p{1.5cm}|p{1.5cm}|p{1.6cm}|} 
 \hline
 & & & \\[-0.5em]
 Solvent & $\mathrm{\lambda^{M-H}(eV)}$ & $\mathrm{\lambda^{MHC}(eV)}$ & $\mathrm{\lambda^{M-H\diamond}(eV)}$ \\[0.3em]
  & & & Ref. [\citenum{boyle2020transient} ] \\
 \hline\hline
 PC& 0.33 & 0.21& 0.34\\
 \hline
 DEC& 0.38& 0.25& 0.34\\
 \hline
 EC:DEC& 0.34& 0.22& 0.3\\
 \hline
 EC:DEC w. 10\% FEC&0.31 & 0.19& 0.28\\
 \hline
\end{tabular}
\caption{The reorganization energy values as obtained from the M-H model, $\mathrm{\lambda^{M-H}}$, the MHC model, $\mathrm{\lambda^{MHC}}$, and values reported by Boyle \textit{et al}., from another implementation of the M-H model.}\label{table:lamfit}
\end{table}

\begin{center}
\begin{figure}[ht]
    \includegraphics[width=\textwidth]{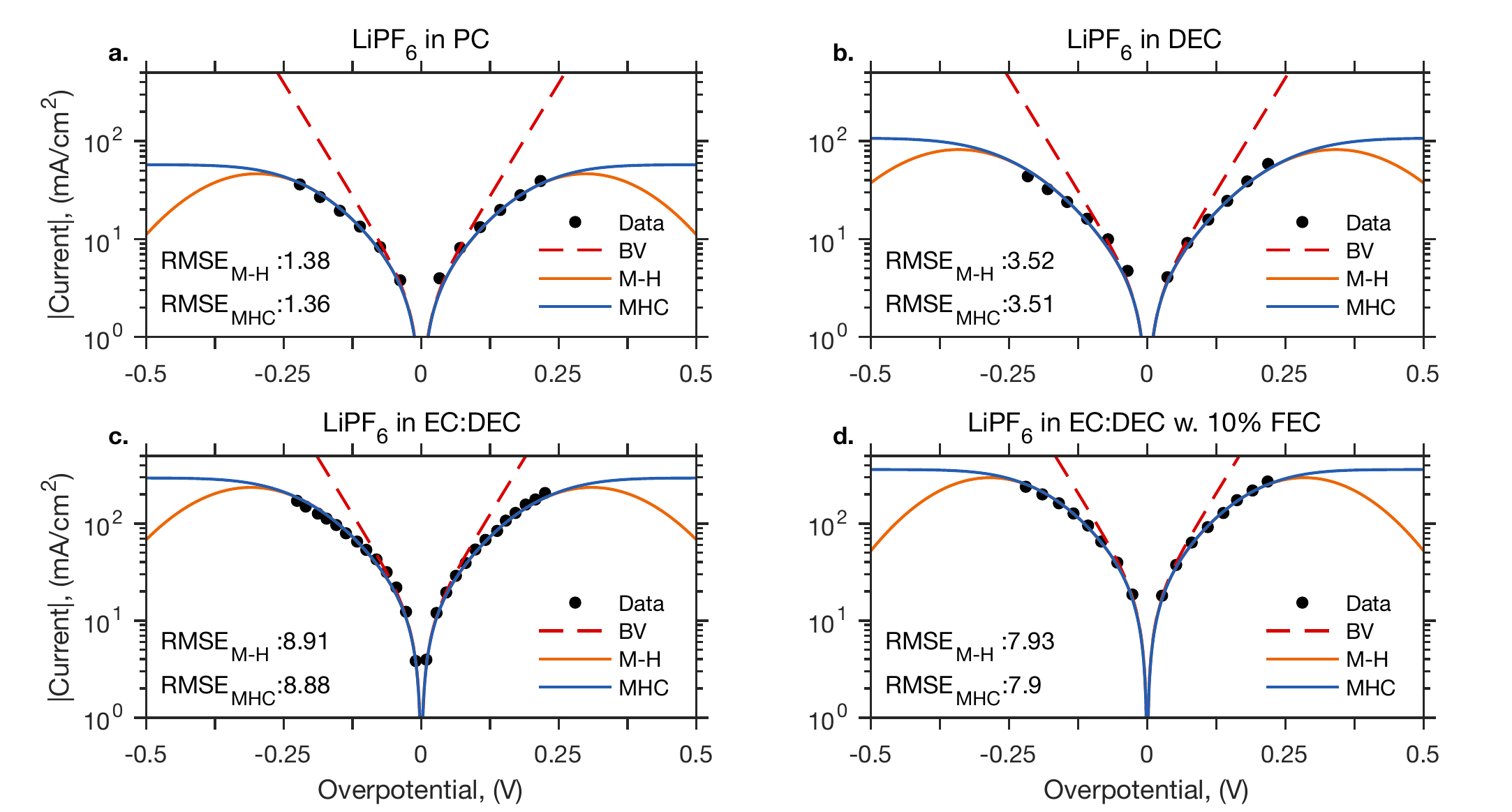}
    \caption{The predictions of current density as provided by the BV model, the M-H model (proposed by Boyle \textit{et. al})\cite{boyle2020transient} and the MHC model for four different solvents, namely, a. PC, b. DEC, c. EC:DEC, d. EC:DEC w. 10\% FEC. The exchange current densities and the reorganization energy values for each of the fits are shown in Tables \ref{table:jofit} and \ref{table:lamfit} respectively.The root mean square error (RMSE) in $\mathrm{mA/cm^2}$ for M-H and MHC models is very similar while the goodness of fit, characterized by $\mathrm{R^\text{2}}$ is the same for both M-H and MHC for each solvent ($\mathrm{R^\text{2}}$ for both M-H and MHC models are 0.997, 0.987, 0.992, and 0.997 for PC, DEC, EC:DEC, and EC:DEC w. 10\% FEC respectively). It is evident that both M-H and MHC models explain the data very well and collapse on each other at low overpotentials, however, at overpotentials greater than an absolute value of 0.25V, the difference in predictions is significant. This is because the M-H model posits an inverted region for current at overpotentials where $\mathrm{\eta>\lambda}$, while the MHC models predicts a plateau. Given that these models are examined in the context of using them in lithium battery models in high power conditions, it is important to use the more appropriate model for overpotentials greater 0.25V. Predictions from the BV model are also shown here for reference, where the BV model uses the same exchange current density of the M-H model. As seen for each solvent, the BV model deviates significantly for overpotentials greater than 0.1V.}
    \label{fig:kinetics}
\end{figure}
\end{center}

\begin{center}
\begin{figure}[ht]
    \centering
    \includegraphics[width=\textwidth]{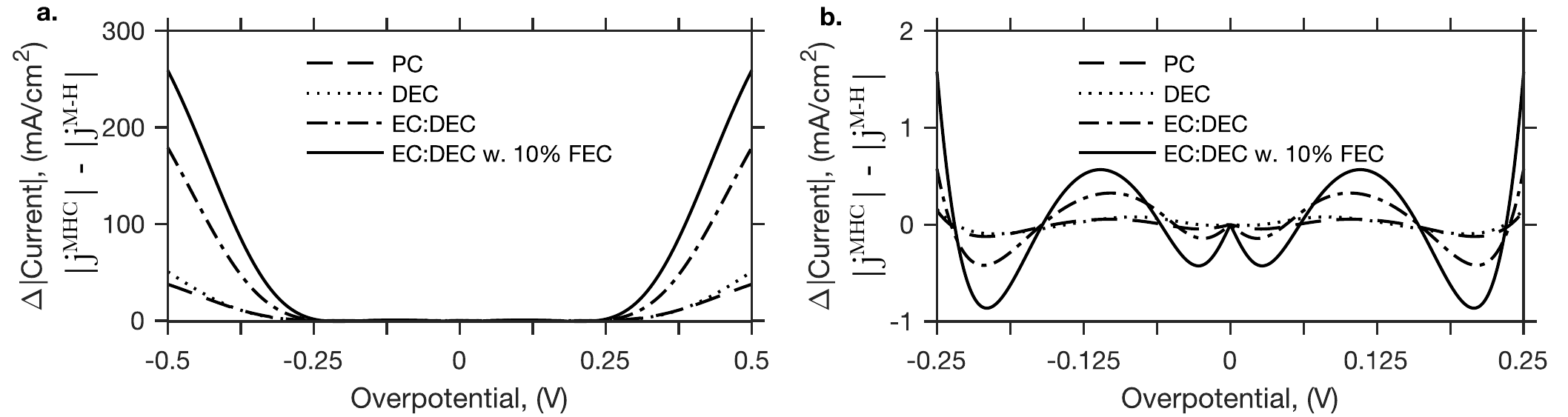}
    \caption{A comparison of the predictions for current density from M-H and MHC models. a. Shows the difference between $\mathrm{j^{MHC}}$ and $\mathrm{j^{M-H}}$ for the potential window between -0.5V and 0.5V. b. The difference between $\mathrm{j^{MHC}}$ and $\mathrm{j^{M-H}}$ for the potential range of -0.25V and 0.25V which are the limits of the overpotential in the dataset. We observe that the difference is about 50 $\mathrm{mA/cm^2}$ for PC and DEC while it goes up to 200 $\mathrm{mA/cm^2}$ for EC:DEC at 0.5V of overpotential and about 250 $\mathrm{mA/cm^2}$ at 0.5V for EC:DEC w. 10\% FEC. Within an overpotential region of 0.25V, the greatest error is about 1.5 $\mathrm{mA/cm^2}$ for EC:DEC w. 10\% FEC. when the absolute value of current density is over 200 $\mathrm{mA/cm^2}$ which is a difference of under 1\%. As explained in Figure \ref{fig:kinetics}, it is clear that both the M-H and MHC models explain the data with almost equal accuracy, with low error and small difference between the two models for overpotential regions within an absolute value of 0.25V, however, the difference in predictions is significant when we consider a slightly higher overpotential window of about 0.5V.}
    \label{fig:kineticserrors}
\end{figure}
\end{center}

The MHC model consistently predicts lower values of reorganization energy for all four solvents, while consistently maintaining similar values for exchange current density with the M-H model. The only difference between the M-H model (implemented using a potential depending transfer coefficient $\mathrm{\alpha}$) and the MHC model (implemented using the uniformly valid closed form approximation in \eqref{mhcapprox}), is the use of Fermi-Dirac statistics to account for distribution of electrons in different energy levels. Hence, the difference is reorganization energies can be attributed to the inclusion of the distribution of electrons over different energy levels. Further, the M-H implementation discussed here and as implemented in Boyle \textit{et al}., is a low overpotential approximation of the MHC model. Hence, it follows that the use of the M-H model should be limited to overpotential regions where $\mathrm{\eta<\lambda}$. Further, Bai \textit{et al}.,\citenum{bai2014charge} have confirmed from first principles calculations that the reorganization energies predicted by the MHC model is more accurate compared to typical Marcus or M-H models. Bai \textit{et al}.,\citenum{bai2014charge} also note that Marcus or M-H models require larger reorganization energies to fit electrode kinetics data because of the inverted region predicted for $\mathrm{\eta>\lambda}$. Between the two models, the MHC rate expression implemented using the closed form approximation in \eqref{mhcapprox} is more appropriate to use within an electrochemical battery model, as shown in discussions pertaining to Figures \ref{fig:kinetics} and \ref{fig:kineticserrors}.

A comparison between the different models and how they explain the transient voltammetry data is shown in Figure \ref{fig:kinetics}. One of the first observations that can be made is the significant deviation of BV model from the data for all solvents. This observation has been noted for several metal electrodeposition and stripping studies,\cite{cogswell2015qualitative,fawcett1989potential} and pointed out for the case of lithium as well.\cite{boyle2020transient} Further, we can see that the M-H model proposed by Boyle \textit{et al}., and the MHC model from this work explain the data with similar accuracy, however, the current density predictions from the two models begin to deviate significantly after about 0.3V of overpotential. The M-H model predicts an inversion of current similar to the inverted region proposed by Marcus theory.\cite{marcus1956theory,marcus1965theory} As noted previously, the M-H implementation shown here and in Boyle \textit{et al}., is a low overpotential approximation of the MHC model while the MHC model implement in this work is a uniformly valid approximation. Hence, we see that the two models are indistinguishable at lower overpotentials. The difference between the models is significant at overpotentials greater than 0.25V and deviation is larger for larger overpotentials.
\onecolumngrid
\twocolumngrid

\onecolumngrid
\twocolumngrid
The deviation of the M-H model from MHC model starts at about 0.25V of overpotential. Figure \ref{fig:kineticserrors}a and \ref{fig:kineticserrors}b show the difference between the absolute values of current from the two kinetic models. Figure \ref{fig:kineticserrors}b shows the deviation within an overpotential region of 0.25V which is the maximum overpotential in the transient voltammetry data.\cite{boyle2020transient} We observe that, for the overpotential region within 0.25V, the maximum deviation of about 1.5 $\mathrm{mA/cm^2}$ is caused for the high exchange current density solvent EC:DEC w. 10\% FEC at 0.25V of overpotential. The difference between the currents from the two kinetics models is much lower for the other three solvents. This differences are very small compared to the absolute value of the current at this potential which is about 250 $\mathrm{mA/cm^2}$. On the other hand, on examining the difference in currents for the overpotential region within 0.5V, we observe that, for EC:DEC w. 10\% FEC, the deviation of the prediction of current from the M-H model is about 250 $\mathrm{mA/cm^2}$. Similarly, for lower exchange current density solvents like PC and DEC, the deviation is about 50 $\mathrm{mA/cm^2}$. Due to the inverted region predicted by the M-H model, the current predictions are always an underestimate at overpotentials greater than 0.25V for the M-H model.

Based on the discussions here, and the results shown in Figures \ref{fig:kinetics} and \ref{fig:kineticserrors}, while the MHC model collapses onto the M-H model at low overpotentials, the deviation of the predictions from the M-H model are substantial at slightly higher overpotentials here. Since these deviations are significant, and that the M-H model is only a low overpotential approximation of the MHC model, the MHC model implemented using the uniformly valid closed form approximation,\cite{zeng2014simple} is better suited to be used with electrochemical models of lithium batteries. The exchange current density fit for both models are identical, however, the reorganization energies are lower by about 0.12eV. Further experiments beyond overpotentials of 0.25V are required to confirm the nature of the Tafel curves for lithium electrodeposition and stripping at higher overpotentials.

\subsection{Pseudo-2D fast-charge model}
Implementing the kinetics models described in Section \ref{methodskinetics} within the P2D model framework described in Section \ref{p2dmodel} demonstrates the importance of accurate charge-transfer kinetics at the application scale. While the differences between the three mechanisms may appear subtle in Figures~\ref{fig:kinetics} and~\ref{fig:kineticserrors}, results predict significant performance differences when employed at the cell level. Accurate model predictions will be critical to inform battery design and control strategies for cases such as XFC in EVs or eVTOL vehicles.

\begin{center}
\begin{figure}[hb]
    \includegraphics[width=0.45\textwidth]{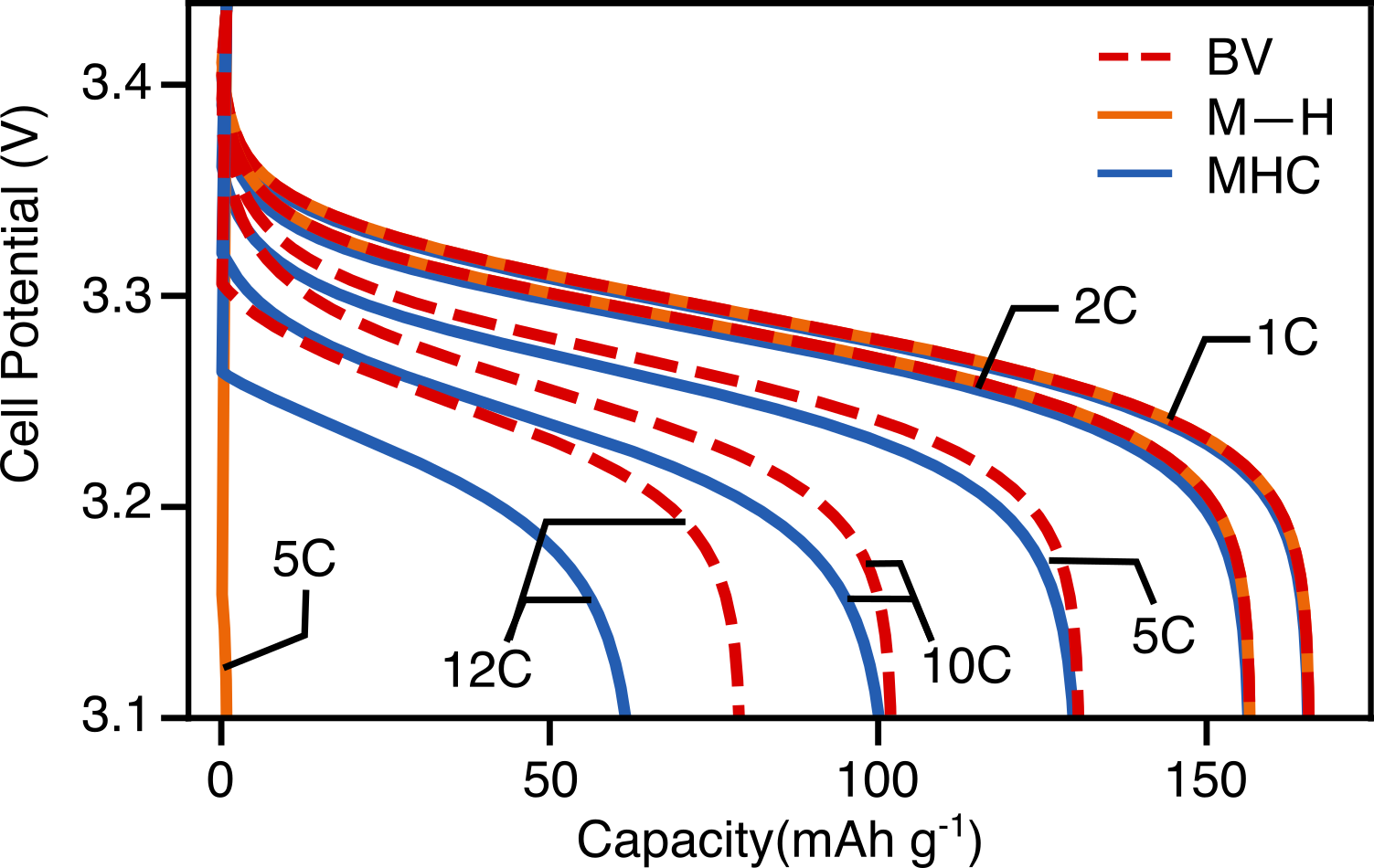}
    \caption{P2D model predictions for XFC discharge with C-rates from 1C to 12C. Comparing results for the three kinetic models demonstrate the importance of accurate charge transfer kinetics at the cell level. Above discharge rates of roughly 2C, we see significant differences between discharge profiles for the Butler-Volmer (BV), Marcus-Hush (M-H), and Marcus-Hush-Chidsey (MHC) models. In particular, the MHC and M-H models predict much higher voltage losses with increasing charge rates, compared to the commonly-used Butler-Volmer form. Moreover, the M-H model is unable to sustain discharge at rates of 5C or higher, due to the so-called `inverted region.'}
    \label{fig:discharge}
\end{figure}
\end{center}

Figure~\ref{fig:discharge} shows predicted discharge curves for a Li--LiFePO$_4$ cell operating at C-rates ranging from 1C to 12C. At moderately high C-rates (1C and 2C), the models are in good agreement throughout discharge. However, for C-rates above 2C, model results diverge for the three models. Particularly, the M-H model is unable to handle discharge currents above 5C. In addition to this, the MHC and BV models diverge at discharge currents of 5C and above. The scale of this divergence increases with increasing C-rates. This agreement at low discharge currents with significant divergence from agreement at high currents is a result of the predictions made in Figures~\ref{fig:kinetics} and~\ref{fig:kineticserrors}. Not only do the voltage curves diverge, but currents over 10C show a significant difference in discharge capacity. 
\begin{center}
\begin{figure}[t]
    \includegraphics[width=0.45\textwidth]{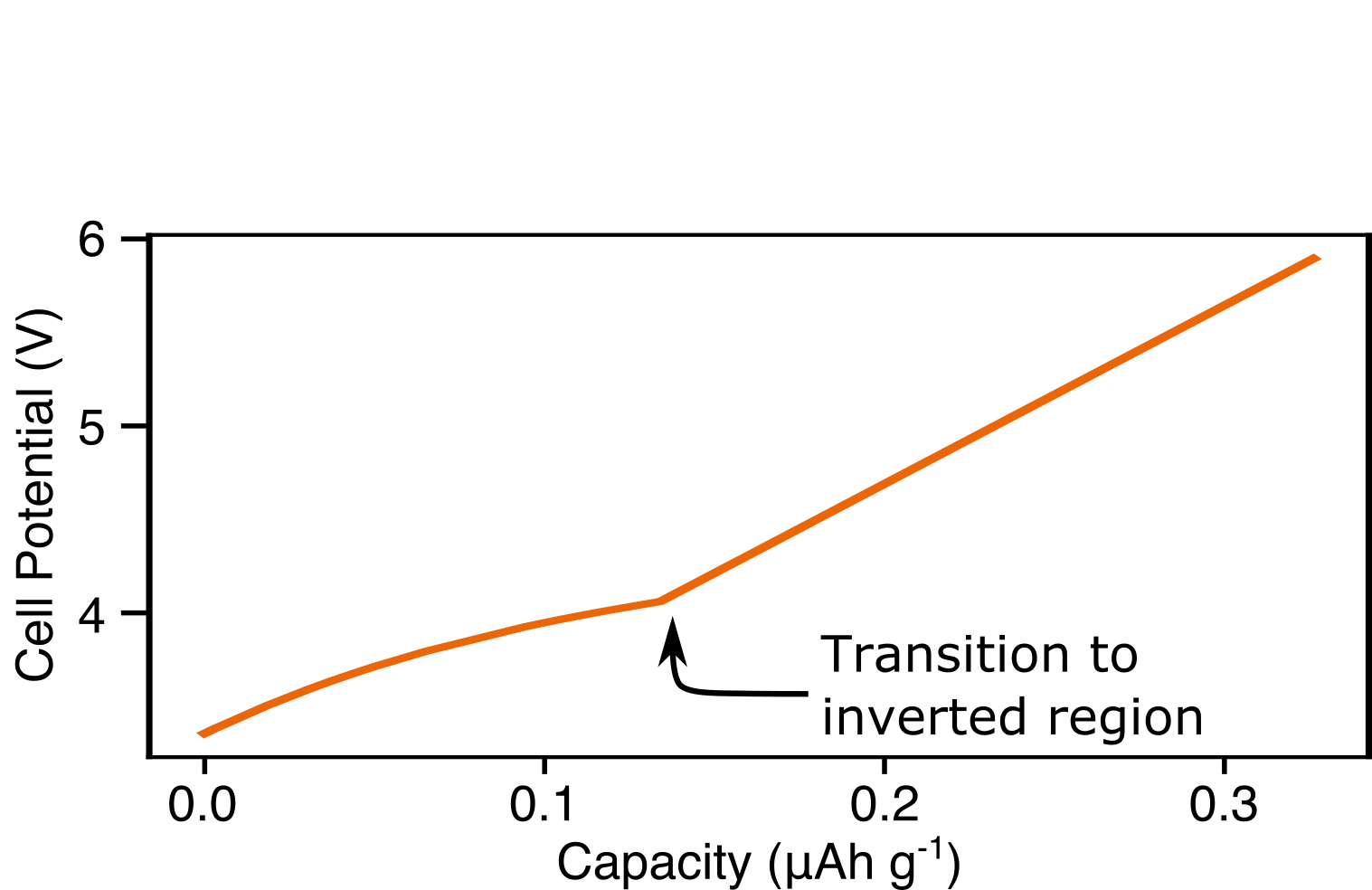}
    \caption{P2D-predicted charging profile with the M-H kinetic model and a charging current of 5C demonstrating the transition from the charge transfer current predicted at low overpotentials to the 'inverted region' at high overpotentials.}
    \label{fig:inverted}
\end{figure}
\end{center}
The M-H model predictions in Figure~\ref{fig:discharge} are explained by the so-called `inverted region' in Figure~\ref{fig:kinetics}, where the current density \emph{decreases} as a function of the overpotential $\eta$, for $\eta$ values significantly greater than the solvent reorganization energy $\lambda$.  This phenomena is shown in greater detail in Figure~\ref{fig:inverted}, which plots the charging voltage for the M-H model at 5C. The voltage initially increases, due to charging of the double layer.  As the resulting Faradaic current density at the cathode-electrolyte interface increases commensurately with $\mathrm{\Delta \Phi_{dl}}$, the difference between the $\mathrm{j_{ext}}$ and the total $\mathrm{j_{Far}}$ decreases, and the slope of the voltage profile also decreases. However, once the overpotentials reach the point at which $\mathrm{j_{Far}}$ is maximized (roughly $\eta = 0.35$ V, in Figure~\ref{fig:kinetics}), further charging of the double layer decreases the Faradaic current, which then leads to unconstrained growth of the double layer. This is observed in Figure~\ref{fig:inverted} at a capacity of roughly 0.14 $\mu$Ah g$^{-1}$.  As $\mathrm{j_{Far}}$ goes to zero with further increases in $\mathrm{\Delta \Phi_{dl}}$, the double layer current is essentially constant at $\mathrm{j_{dl} = j_{ext}}$, as reflected in the linear growth rate for the cell potential at discharge capacities greater than 0.14 $\mu$Ah g$^{-1}$ (note that capacity is proportional to time, for galvanostatic discharge). The absence of the inverted region in the systems studied here (due to the presence of valence-band electrons) enables sustained charge-discharge at high C-rates, and cautions against the use of M-H kinetics for XFC predictions.

\begin{center}
\begin{figure}[hb]
    \includegraphics[width=0.45\textwidth]{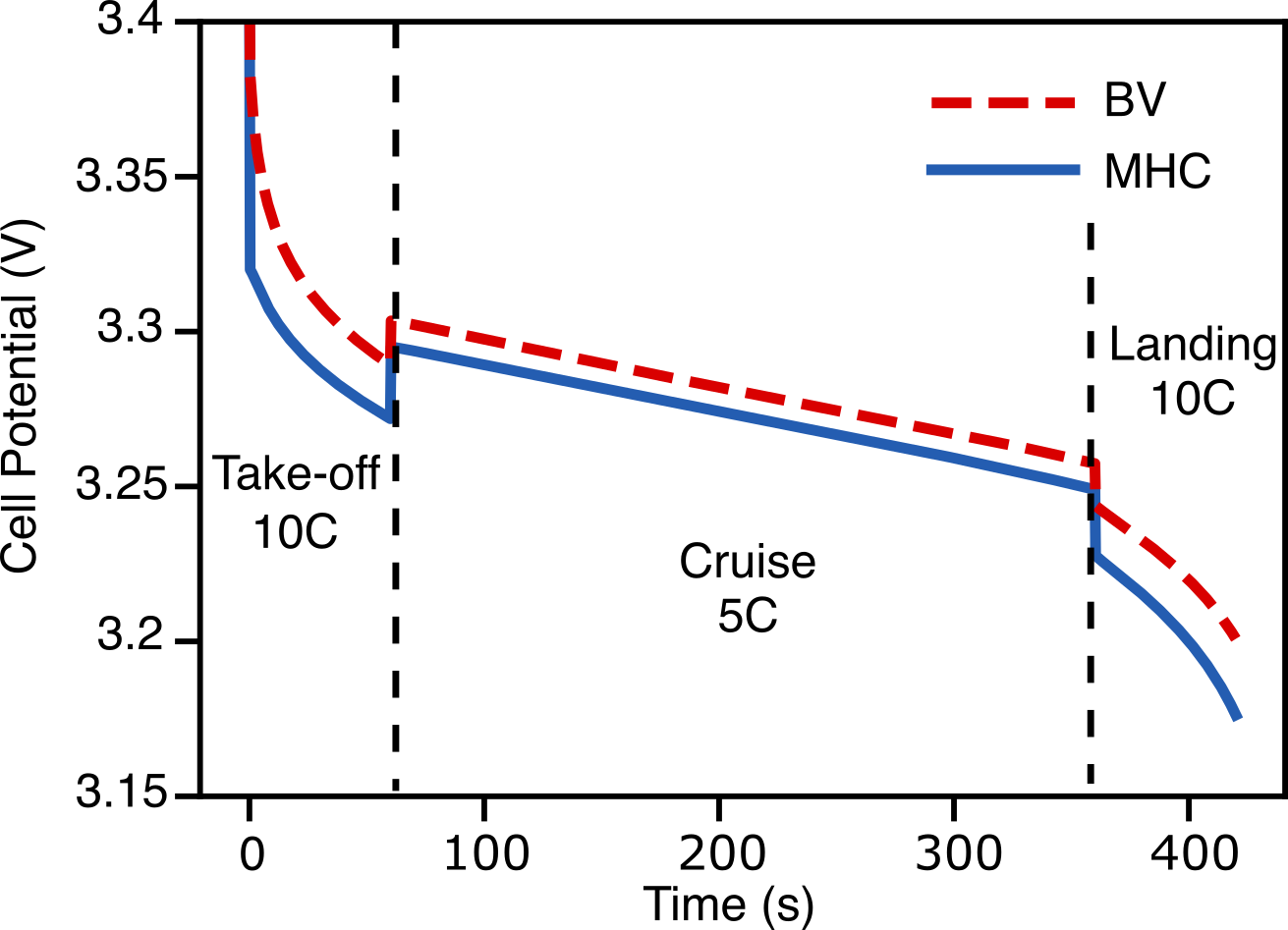}
    \caption{Predicted discharge curve of a time-dependent current simulating the changing load during takeoff, flight, and landing that an eVTOL might experience. The MHC and BV kinetics models predict different voltage and power throughout the flight time. This is particularly evident during the `Take-off' and `Landing' segments. However, due to the relatively lower current demand during the 'Cruise' segment, the disparity between the kinetic model predictions decreases and remains roughly constant for the duration of the flight period.}
    \label{fig:VTOL}
\end{figure}
\end{center}

A comparison of the MHC and the BV kinetic models' predictions in the context of an eVTOL flight is shown in Figure~\ref{fig:VTOL}, demonstrating the importance of the kinetics model used when predicting performance and battery design needs in such cases. Due to the high current demand during takeoff for an eVTOL, as predicted by the results in Figure~\ref{fig:discharge}, the MHC model predicts a lower power for the same discharge time than the BV model, on the order of mW cm$^{-2}$. In addition to different predictions in power when the two kinetics models are employed, as shown in Figure~\ref{fig:discharge}, when discharging at a current over 10C, the predicted discharge capacity is different, which can have non-negligible impacts on flight duration and battery sizing. Far from a merely academic consideration, the detailed charge transfer kinetics investigated here have meaningful implications on battery design and sizing for emerging fast-charge, high-power applications such as eVTOL.

\section{Conclusions}
In this work, we used transient voltammetry data for lithium electrodeposition and stripping from Boyle \textit{et al}., to explore the compatibility of different kinetic theories to the data. The kinetic theories are compared in the context of developing better kinetic models for use in electrochemical performance simulations of lithium batteries. The widely used Butler-Volmer theory is found to be incompatible with the voltammetry data at high rates. We compare the use of the M-H kinetic model, proposed by Boyle \textit{et al}., with the MHC model to explain the voltammetry data for \ce{LiPF6} in four solvents, PC, DEC, EC:DEC, and EC:DEC w. 10\% FEC. We find that both the M-H model and the MHC model fit the data with equal accuracy. The exchange current densities fit for each of the solvents are similar for the M-H and the MHC models. 

We find that the reorganization energies predicted in this work for the M-H model match the reorganization energies predicted by Boyle \textit{et al}. However, we find that the MHC model consistently predicts a (0.12$\pm0.1$eV) lower reorganization energy. We attribute this difference to the incorporation of the distribution of electrons at different energy levels within the MHC model, which the M-H model ignores. This is also consistent other studies\cite{bai2014charge} where it was found that the reorganization energy predicted by the MHC model was more accurate and that Marcus and M-H models required higher reorganization energies to fit Tafel curves. Given that the MHC predicted reorganization energies are more accurate than Marcus or M-H predictions for other materials, we believe that this could hold for the lithium/solvent systems studied here as well. We also note that the M-H model proposed by Boyle \textit{et al}., is a low overpotential approximation of the MHC model, and hence the two models are indistinguishable at low overpotentials.  At higher overpotentials, we observe significant deviations between the predictions of the M-H and MHC models.  Thus, we propose the use of the MHC model implemented using a uniformly valid closed form approximation within battery models for describing the kinetics at lithium electrodes. 

This recommendation is supported by P2D battery simulations of extreme fast charge of a lithium metal anode paired with a carbon-LiFePO$_4$ composite cathode. The M-H kinetics predict that the battery cannot sustain charge-discharge at rates of 5C or greater.  The Butler-Volmer and MHC kinetics, on the other hand, predict relatively similar performance up through discharge at 5C, but diverge significantly at 10C and above.  Using an example eVTOL mission, we demonstrate that the two models give significant differences in delivered power during the mission.

While the Butler-Volmer and M-H models offer the advantage of computational simplicity, closed-form approaches for MHC have been published and can be readily implemented for validation against experimental data.~\cite{zeng2014simple} To this end, we have published the model and thermo-kinetic modeling tools used here as open-source software, to facilitate and accelerate the adoption of MHC kinetics more broadly throughout the field.

\section{Supplementary Material}
See supplementary material for more visualizations of the fitting of kinetic models.
\begin{acknowledgments}
S.S. and V.V gratefully acknowledge funding support from Aurora Flight Sciences.  D.M.K. and S.C.D. gratefully acknowledge funding support from the National Science Foundation under grant number 1903440, Program Officer Raymond Adomaitis.  S.C.D and D.M.K. also wish to thank Prof. Wolfgang Bessler and Lutz Schiffer, Offenburg University of Applied Sciences, for insightful conversations about charge-transfer software implementations. The authors thank the Battery Modeling Webinar Series (BMWS) community for nucleating this collaboration and Battery Modeling Slack workspace for helping manage this collaboration.
\end{acknowledgments}
\clearpage
\bibliography{refs}

\clearpage
\includepdf[pages={1}]{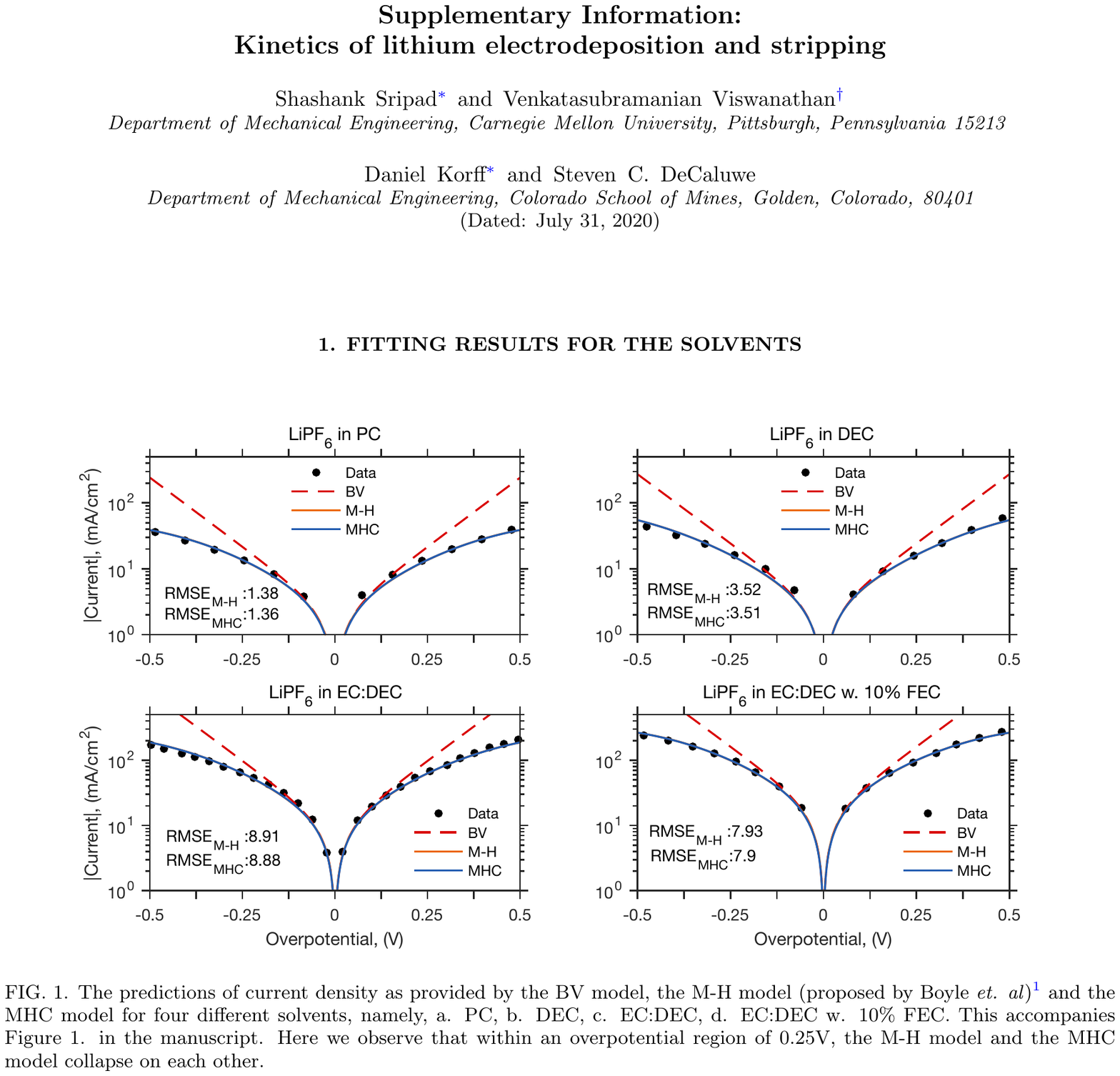} 
\clearpage
\includepdf[pages={2}]{SI}

\end{document}